\documentclass[aps,pra,superscriptaddress,floatfix,showpacs,10pt,twocolumn]{revtex4}
\usepackage{amsmath}
\usepackage{graphicx}%
\usepackage{amsfonts}
\usepackage{amssymb}

\usepackage[T1]{fontenc}
\usepackage[latin2]{inputenc}
\usepackage{latexsym}
\usepackage{bbm}
\usepackage{color}

\begin{document}
\title{Sudden death of distillability in qutrit-qutrit systems}
\author{Wei Song}
\email{wsong1@mail.ustc.edu.cn} \affiliation{Laboratory of Quantum
Information Technology, ICMP and SPTE, South China Normal
University, Guangzhou 510006, China}
\author{Lin Chen}
\email{cqtcl@nus.edu.sg(Corresponding~Author)} \affiliation{Centre
for Quantum Technologies, National University of Singapore, 3
Science Drive 2, Singapore 117542}
\author{Shi-Liang Zhu}
\affiliation{Laboratory of Quantum Information Technology, ICMP and
 SPTE, South China Normal University, Guangzhou 510006, China}

\date{\today}
\date{2 Jul 2009}


\begin{abstract}
We introduce the concept of distillability sudden death, i.e., free
entangled states can evolve into non-distillable (bound entangled or
separable) states in finite time under local noise. We describe the
phenomenon through a specific model of local dephasing noise and
compare the behavior of states in terms of the Bures fidelity. Then
we propose a few methods to avoid distillability sudden death of
states under (general) local dephasing noise, so that free entangled
states can be robust against decoherence. Moreover, we find that
bound entangled states are unstable in the limit of infinite time.

\end{abstract}
\maketitle

\section{Introduction}

Entanglement is not only a remarkable feature which distinguishes
the quantum world from the classical one but also a key resource to
realize high-speed quantum computation and high-security quantum
communication \cite{Neilsen:2000}. In realistic quantum-information
processing, entanglement usually needs to be prepared or distributed
beforehand among different remote locations. However, in the process
of entanglement distribution the quantum systems are not isolated
and each system will unavoidably interact with the environment. This
leads to local decoherence which will degrade the entanglement of
the shared states. It is thus of fundamental importance to study the
entanglement properties under the influence of the local
decoherence. In this context, Yu and Eberly \cite{Yu:2004}
investigated the time evolution of entanglement of a bipartite qubit
system undergoing various modes of decoherence. Remarkably, they
found that, although it takes infinite time to complete the
decoherence locally, the global entanglement may vanish in finite
time. The phenomenon of finite-time disentanglement, also named
entanglement sudden death(ESD), unveils a fundamental difference
between the global behavior of an entangled system and the local
behavior of its constituents under the effect of local decoherence.
Clearly, ESD puts an limit on the applicability of entangled states
in the practical quantum information processing.

Initially, Yu and Eberly reported the ESD for two-qubit entangled
states, but this effect is not limited to such case. Further
investigations in a wider context including higher dimensional
Hilbert spaces have been made by various groups
\cite{Yonac:2006,Cui:2007,Yu:2007,Lastra:2007,Zhang:2007,Vaglica:2007,Bellomo:2007,Ikram:2007,Qasimi:2008,Ficek:2006,Sainz:2008,Ann:2007,Rau:2008,Huang:2007,Li:2009,Shan:2008,Fei:2009}.
There are also a number of studies looking at ESD in more
complicated systems using other entanglement measures
\cite{Carvalho:2007,Annb:2007,Sainz:2007,Lopez:2008,Aolita:2008,Man:2008,An:2009,Liu:2009},
and an attempt to give a geometric interpretation of the phenomenon
has also been made \cite{Cunha:2007}. In addition, experimental
evidences of ESD have been reported for optical setups
\cite{Almeida:2007} and atomic ensembles \cite{Laurat:2007}.

However, all previous studies omitted an important fact that high
dimensional bipartite entangled states can be divided into two
classes \cite{Horodecki:1998}. One is free, which means that the
state can be distilled under local operations and classical
communication (LOCC); the other is bound, which means that no LOCC
strategy is able to extract pure-state entanglement from the state
even if many copies are available. Bound entanglement(BE) cannot be
used alone for quantum information processing, and irreversibility
occurs in asymptotic manipulations of entanglement for all BE
states\cite{Yang:2005}. Since it was constructed from a pure
mathematical point of view, we may ask whether BE can appear in
physically relevant quantum systems naturally. Very recently a few
works have addressed this question \cite{Toth:2007}. Their results
suggest that different many-body models present thermal bound
entangled states.

In this paper, we investigate the problem from a very different
viewpoint, i.e., in the present of local decoherence, which is one
of the dominant noises during the distribution of entanglement.
Analogous to the definition of ESD, if an initial free entangled
state becomes non-distillable in finite time under the influence of
local decoherence, then we say that it undergoes distillability
sudden death(DSD). Note that when a free entangled state loses its
distillability at a specific time, it may still be entangled since
we cannot exclude the existence of BE. So far, it is not clear
whether bipartite bound entangled states can be created from free
entangled states under local decoherence process naturally. The
first aim of this paper is to show that such a process indeed exists
through an explicit qutrit-qutrit example.

Afterwards we propose the DSD-free state, which has entanglement
robust against local decoherence. Such entangled states are thus
useful resources for practical quantum-information processing. So
the second aim is to address the DSD-free state. We develop a few
systematic approaches to build DSD-free states. Finally we will show
that different from free entangled and separable states, no PPT
bound entangled states exist in the infinite time limit.

The paper is organized as follows. In Sec. II, the idea of DSD is
presented. In particular, we show the phenomenon of DSD through a
specific qutrit-qutrit example. In addition, we show that one can
avoid the sudden death of distillability by performing a simple
local unitary operation on the initial state. Furthermore, we
compare the behaviors of states affected by decoherence in terms of
the Bures fidelity. In Sec. III, we develop some methods to build
DSD-free states which can protect free entanglement under general
local dephasing noise. We also show that no PPT entangled states can
exist in the limit of infinite time. Finally, in Sec. IV we discuss
some open questions and also give a summary of our results.

\section{Qutrit-Qutrit DSD states under local decoherence}

Before discussing dynamical process of entanglement, we briefly
review how to characterize bound entangled states. It was proven in
\cite{Horodecki:1998} that a quantum state with a positive partial
transposition(PPT) is non-distillable under LOCC. Therefore PPT
entangled states must be bound entangled states \cite{Song:2009}. To
verify them, one can use the so-called realignment criterion
(cross-norm criterion) \cite{Rudolph:2005}. The definition of
realignment on the density matrix is given by $ \left( {\rho ^R }
\right)_{ij,kl}  = \rho _{ik,jl}$. A separable state $\rho$ always
satisfies ${\left\| {\rho ^R } \right\| \le 1}$. For a PPT state
$\rho$, the positive value of the quantity ${\left\| {\rho ^R }
\right\| - 1}$ can hence verify that it is a bound entangled state.

\begin{figure}[ptb]
\includegraphics[scale=0.55,angle=0]{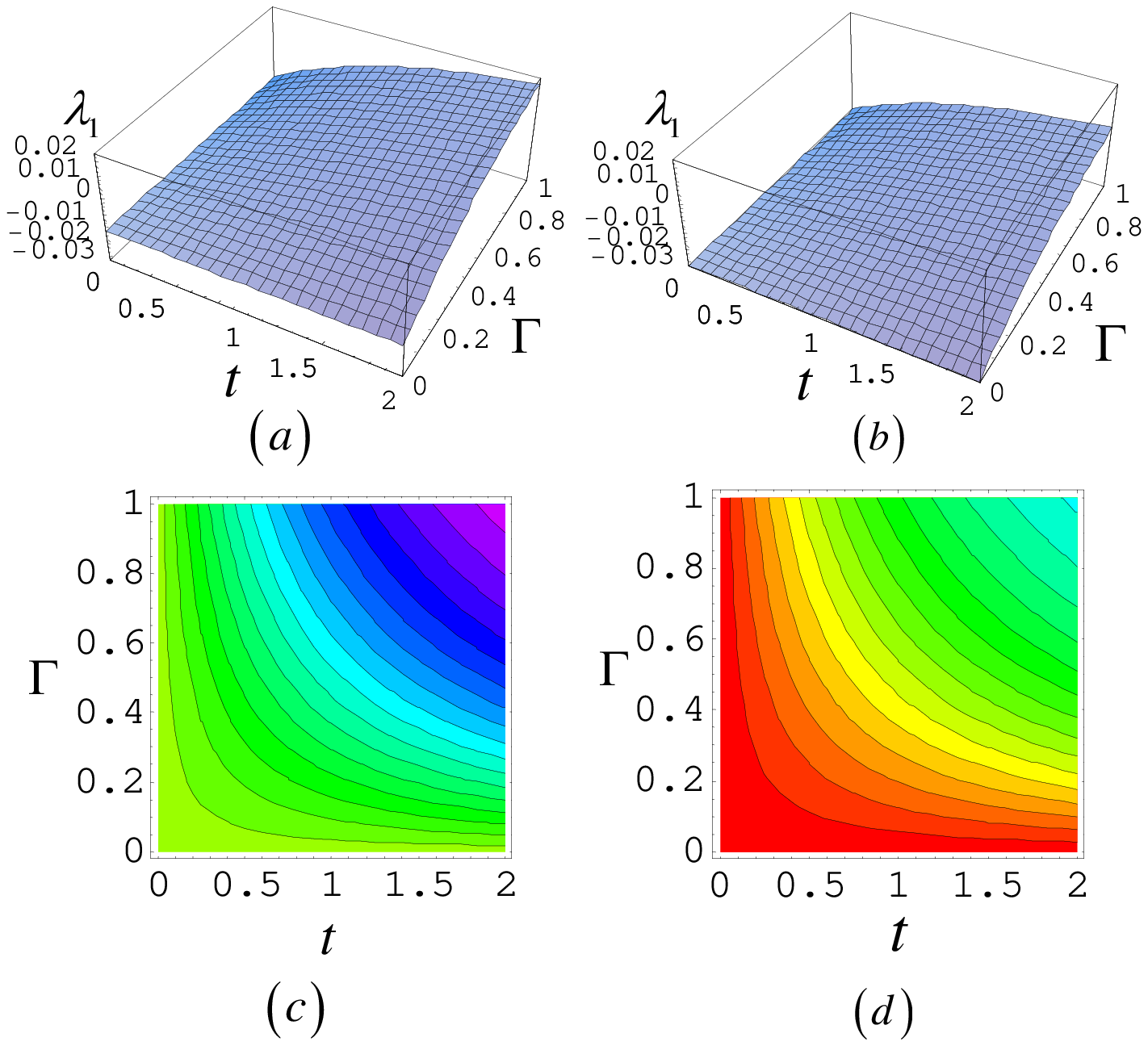}\caption{
(Color online). The eigenvalue $ \lambda _1$ as a function of time
$t$ and dephasing rate $\Gamma$ for two different cases: (a)$ \alpha
= 4.5$; (b)$ \alpha  = 4.9$. Figures (c) and (d) correspond to the
contour plots of the upper one, respectively.}
\label{fig1}%
\end{figure}

The system we study consists of two noninteracting qutrits in two
independent local environments, each coupling to one of the qutrits.
Here the qutrit is a three-dimensional state composed of the
computational basis $  \left| 0 \right\rangle$, $ \left| 1
\right\rangle$, and $ \left| 2 \right\rangle$. Besides, we take the
weak local dephasing noise to model the environment, such noise is
indeed one of the main decoherence sources in solid-state systems.

The general time-evolved
density matrix expressible in the operator-sum decomposition is the
completely positive trace preserving map \cite{Kraus:1983} $ \rho
\left( t \right) = \varepsilon \left( {\rho \left( 0 \right)}
\right) = \sum\nolimits_\mu  {K_\mu ^\dag \left( t \right)\rho
\left( 0 \right)} K_\mu  \left( t \right)$. The operators $ \left\{
{K_\mu \left( t \right)} \right\}$ representing the influence of
statistical noise  satisfy the completeness condition $
\sum\nolimits_\mu {K_\mu  \left( t \right)K_\mu ^\dag \left( t
\right)}  = \mathbb{I}$ which guarantees that the evolution is
trace-preserving\cite{Neilsen:2000}. In our model, the operators are
of the form $K_{\mu}(t) = D_{j}(t)E_{i}(t)$ such that
\begin{eqnarray}
\rho\left(t\right) = \mathcal{E}\left(\rho\left(0\right)\right) =
\sum_{i,j = 1}^{2}
D_{j}^{\dagger}\left(t\right)E_{i}^{\dagger}\left(t\right)
\rho\left(0\right)
E_{i}\left(t\right)D_{j}\left(t\right)\label{krausGeneral}.
\end{eqnarray}
Here, $E_{i}(t)$ and $D_{j}(t)$ correspond to local dephasing noise
components acting on the first and second qutrits, respectively, and
both operators satisfy the completeness conditions. For simplicity,
we first take these to be of the specific forms \cite{Ann:2007}
\begin{eqnarray*}
E_{1}(t) &=& {\rm diag}(1,\gamma_{\rm A},\gamma_{\rm A}) \otimes
\mathbb{I}_{3} \ ,\\ E_{2}(t)  &=&
{\rm diag}(0,\omega_{\rm A},\omega_{\rm A}) \otimes \mathbb{I}_{3} \ ,\\
D_{1}(t) &=& \mathbb{I}_{3} \otimes {\rm diag}(1,\gamma_{\rm
B},\gamma_{\rm B}) \ ,\\ D_{2}(t)  &=& \mathbb{I}_{3} \otimes {\rm
diag}(0,\omega_{\rm B},\omega_{\rm B})\ ,
\end{eqnarray*}
where $\mathbb{I}_{3}$ is the $3 \times 3$ identity matrix,
$\gamma_{\rm A}\left(t\right) = e^{-\Gamma_{{\rm A}}t/2}, \
\gamma_{\rm B}\left(t\right) = e^{-\Gamma_{{\rm B}}t/2}, \
\omega_{\rm A}\left(t\right) = \sqrt{1-\gamma_{{\rm A}}^{2}(t)},\
{\rm and} \ \omega_{\rm B}\left(t\right) = \sqrt{1-\gamma_{{\rm
B}}^{2}(t)}.$

For concreteness, we illustrate our ideas by considering the
following qutrit-qutrit state,
\begin{eqnarray}
 \rho \left( 0 \right) &=& \frac{2}{{21}}\left( {\left| {01} \right\rangle  + \left| {10} \right\rangle  + \left| {22} \right\rangle } \right)\left( {\left\langle {01} \right| + \left\langle {10} \right| +
{\left\langle {22} \right|}
} \right) \notag\\
  &+& \frac{\alpha }{{21}}\left( {\left| {00} \right\rangle \left\langle {00} \right| + \left| {12} \right\rangle \left\langle {12} \right| + \left| {21} \right\rangle \left\langle {21} \right|} \right) \notag\\
  &+& \frac{{5 - \alpha }}{{21}}\left( {\left| {11} \right\rangle \left\langle {11} \right| + \left| {20} \right\rangle \left\langle {20} \right| + \left| {02} \right\rangle \left\langle {02} \right|}
\right)
\end{eqnarray}
\noindent with $ 4 < \alpha  \le 5$. In fact, it is straightforward
to prove that the state $ \rho \left( 0 \right)$ is free entangled.
By using the local operation $\left( {\left| 0 \right\rangle
\left\langle 0 \right| + \left| 1 \right\rangle \left\langle 1
\right|} \right) \otimes \left( {\left| 0 \right\rangle \left\langle
0 \right| + \left| 1 \right\rangle \left\langle 1 \right|} \right)$,
one can converts $ \rho \left( 0 \right)$ into a $2 \otimes 2$
entangled state, which is afterwards distillable by virtue of the
BBPSSW-Horodecki protocol \cite{Horodecki:1997}. So the state $ \rho
\left( 0 \right)$ is a free entangled state and we take it as the
initial state under the local dephasing noise in Eq. (1). This
evolution can be calculated analytically, i.e., at time $t$ the
two-qutrit density operator $\rho(t)$ reads

\begin{equation}
\left( {\begin{array}{*{20}c}
   {\frac{\alpha }{{21}}} & 0 & 0 & 0 & 0 & 0 & 0 & 0 & 0  \\
   0 & {\frac{2}{{21}}} & 0 & {\frac{2}{{21}}\gamma _A \gamma _B } & 0 & 0 & 0 & 0 & {\frac{2}{{21}}\gamma _A }  \\
   0 & 0 & {\frac{{5 - \alpha }}{{21}}} & 0 & 0 & 0 & 0 & 0 & 0  \\
   0 & {\frac{2}{{21}}\gamma _A \gamma _B } & 0 & {\frac{2}{{21}}} & 0 & 0 & 0 & 0 & {\frac{2}{{21}}\gamma _B }  \\
   0 & 0 & 0 & 0 & {\frac{{5 - \alpha }}{{21}}} & 0 & 0 & 0 & 0  \\
   0 & 0 & 0 & 0 & 0 & {\frac{\alpha }{{21}}} & 0 & 0 & 0  \\
   0 & 0 & 0 & 0 & 0 & 0 & {\frac{{5 - \alpha }}{{21}}} & 0 & 0  \\
   0 & 0 & 0 & 0 & 0 & 0 & 0 & {\frac{\alpha }{{21}}} & 0  \\
   0 & {\frac{2}{{21}}\gamma _A } & 0 & {\frac{2}{{21}}\gamma _B } & 0 & 0 & 0 & 0 & {\frac{2}{{21}}}  \\
\end{array}} \right),
\end{equation}

\noindent where the basis are spanned by $ \left\{ {\left|
{{\rm{00}}} \right\rangle ,\left| {{\rm{01}}} \right\rangle ,\left|
{{\rm{02}}} \right\rangle ,\left| {{\rm{10}}} \right\rangle ,\left|
{{\rm{11}}} \right\rangle ,\left| {{\rm{12}}} \right\rangle ,\left|
{{\rm{20}}} \right\rangle ,\left| {{\rm{21}}} \right\rangle ,\left|
{{\rm{22}}} \right\rangle } \right\}$.

Let us analyze the matrix carefully. There are three eigenvalues of
the partial transposition of the state $\rho \left(t \right)$ which
could be negative, namely $ \lambda _1  = f\left( {\Gamma _A }
\right),\lambda _2  = f\left( {\Gamma _B } \right),$ and $\lambda _3
= f\left( {\Gamma _A  + \Gamma _B } \right)$, where $ f\left(
\lambda  \right) = \frac{{e^{ - \lambda t} }}{{882}}\left(
{105e^{\lambda t}  - \sqrt {11025e^{2\lambda t}  + 1764e^{\lambda t}
\left( {4 - 5\alpha e^{\lambda t}  + \alpha ^2 e^{\lambda t} }
\right)} } \right) $. For simplicity, we choose the local asymptotic
dephasing rates $\Gamma_{\rm A} = \Gamma_{\rm B} = \Gamma$, and thus
$ \lambda _1  = \lambda _2 < \lambda _3$ in the following arguments.
In order to have a vivid illustration, in Fig. 1 we plot $\lambda
_1$ as a function of t and $\Gamma$ with specific $\alpha=4.5$ and
$\alpha=4.9$, respectively. Fig.2(a) shows the value of $\lambda _1$
versus $t$ for different decoherence rates with $\alpha=4.5$, and we
can see that the eigenvalue of the partial transposition of the
state $\rho(t)$ will always arrive at a positive value in finite
time. For example, if we choose $\Gamma=1$, the density matrix
$\rho(t)$  will become a PPT state after time $t \approx 0.58$ in
Fig.2(a). Analytically, the time at which $\rho(t)$ becomes a PPT
state is $ t_d = \frac{1}{{\Gamma }}\ln \frac{4}{{\alpha \left( {5 -
\alpha } \right)}}$.

Next, we use realignment to verify the BE in this evolution. To this
end, we need to compute the quantity $ \left\| {\rho(t) ^R }
\right\|-1$, and it is given by $ \frac{2}{{21}}e^{ - \Gamma t}
\left( {2 + 4e^{\frac{1}{2}\Gamma t} + \left( { - 7 + \sqrt {19 -
15\alpha  + 3\alpha ^2 } } \right)e^{\Gamma t} } \right)$. In order
to make a comparison, we plot $ \left\| {\rho \left( t \right)^R }
\right\| - 1$ as a function of time $t$ by choosing $\alpha  = 4.5$
and $\Gamma  = 1$ in Fig.2(b). We can see that if $ t <0.84$, the
value of $ \left\| {\rho \left( t \right)^R } \right\| - 1$ is
always positive, which indicates that in the range $ 0.58 < t
<0.84$, the two-qutrit system is a bound entanglement state. Thus we
have shown that free initial entangled states can evolve into
non-distillable (bound)entangled states in finite time under the
local external asymtotic dephasing noise.

\begin{figure}[ptb]
\includegraphics[scale=0.6,angle=0]{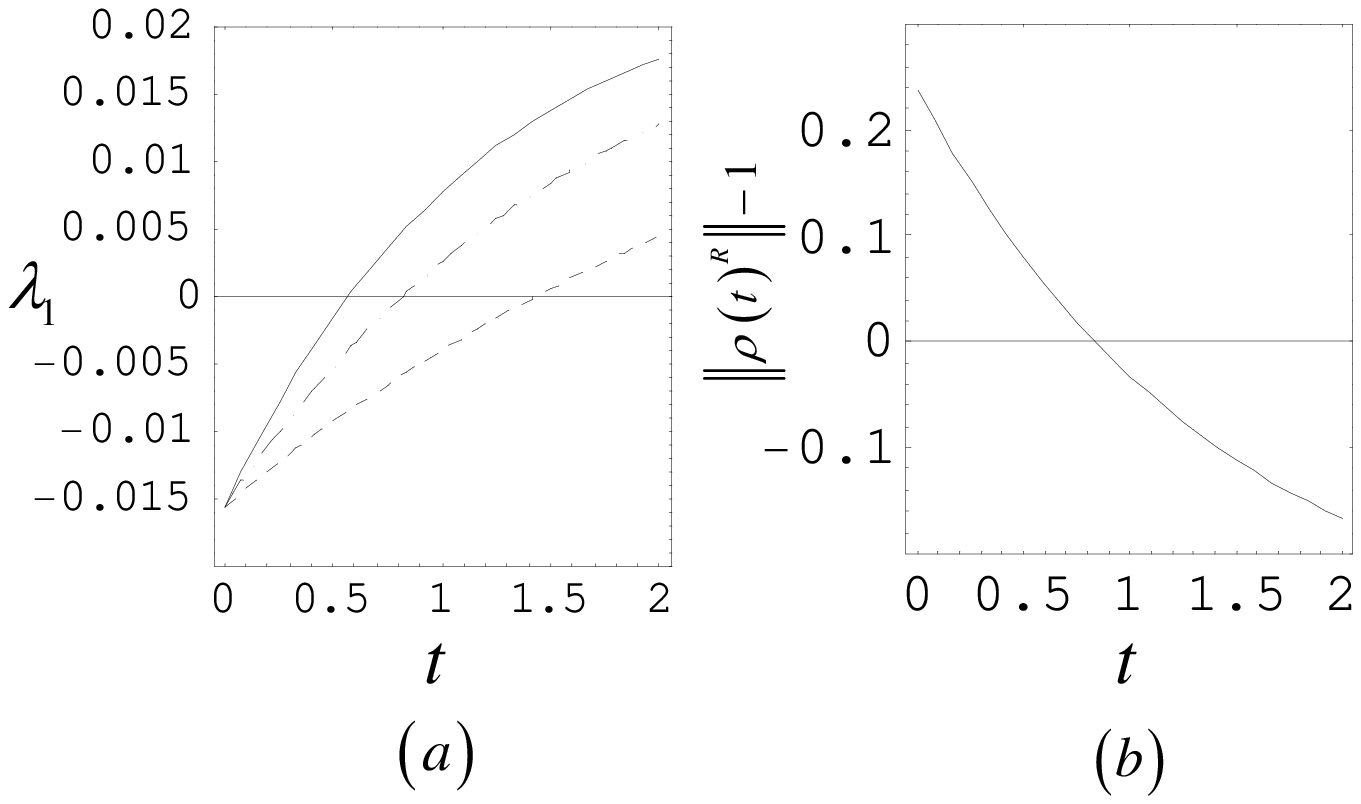}\caption{(a)The eigenvalue $\lambda _1$ versus time $t$ for different dephasing
rates. The solid, dash-dotted, and dashed lines correspond to
$\Gamma = 1,\Gamma = 0.7$, and $\Gamma  = 0.4$, respectively. (b)
The quantity $\left\| {\rho \left( t \right)^R } \right\| - 1$
versus time $t$ for  $\Gamma  = 1$. The other parameter
$\alpha=4.5$.}
\label{fig2}%
\end{figure}

Besides we find that with $\alpha = 4.5$ and $\Gamma  = 1$, the
state $\rho(t)$ will be separable when $t>1.39$. It could be proved
by extracting three "$2\times2$" density operators, which are
respectively spanned by
$\{\left|00\right\rangle,\left|01\right\rangle,\left|10\right\rangle,\left|11\right\rangle\}$,
$\{\left|01\right\rangle,\left|02\right\rangle,\left|21\right\rangle,\left|22\right\rangle\}$
and
$\{\left|10\right\rangle,\left|12\right\rangle,\left|20\right\rangle,\left|22\right\rangle\}$,
from the state $\rho(t)$. Then one can find out that the condition
$t>1.39$ makes the three states separable. As the state $\rho(t)$ is
actually the linear combination of the above three states and some
product states, $\rho(t)$ also becomes separable when $t>1.39$.
Nevertheless, there is still a small window $t\in[0.84,1.39]$ in
which the state $\rho(t)$ has merely PPT and we don't know whether
it's entangled or separable.

To summarize the example, start from the initial state $\rho(0)$ in
$ 4 < \alpha  \le 5$ the qutrit-qutrit state $\rho(t)$ will
experience three phases under the local dephasing noise in Eq.(1):
it evolves from an initial distillable entangled state to a PPT
entangled state, and then changes to be a separable state and
finally remains in this phase for even.
In other word, $\rho(t)$ first experiences DSD and then ESD under
local decoherence.

In contrast with the above example, we briefly consider the initial
state $\rho(0)$ for $ 3 < \alpha  \le 4$. In this case, it is a PPT
entangled state \cite{Horodecki:1999}, one can verify that the state
$\rho(t)$ will finally become separable in finite time under the
local dephasing noise in Eq.(1); that is, $\rho(t)$ only experiences
ESD. Moreover, we can easily verify the time-domain factorization
relation as follows,

\begin{equation}
S_{t_1  + t_2 }  = S_{t_1 } S_{t_2 },
\end{equation}

\noindent where $t_1 ,t_2  \ge 0$, and $S$ denotes the noise in
Eq.(1). Because the entanglement cannot increase under the local
operations, we conclude that if a state becomes separable at a
specific time in Eq.(1), it must remain separable in all subsequent
time. More generally, if an entangled state evolves into a PPT state
(either entangled or separable) at a specific time in Eq.(1), it
always has PPT for all subsequent time since the local operation
cannot change PPT.

Of course, such disappearance of distillability in finite time can
seriously affect the application of entanglement in quantum
information tasks. An important question arises naturally: given a
free initial entangled state, does there exist a suitable
intervention that may alter the final fate of DSD? In the following
we show that one can realize this aim by merely performing a simple
local unitary operation on the initial state in the above example.
Let the initial local operation be $ U = \mathbb{I}_3 \otimes
\lambda$, with $ \lambda  = \left| 0 \right\rangle \left\langle 1
\right| + \left| 1 \right\rangle \left\langle 0 \right| + \left| 2
\right\rangle \left\langle 2 \right| $. Then, the transformed state
is $ \rho '\left( 0 \right) = U\rho \left( 0
\right)U^\dag=\frac{2}{7}P_ +   + \frac{\alpha }{7}\rho _ +   +
\frac{{5 - \alpha }}{7}\rho _ -  ,$ where the projectors are $P_ + =
\left| {\Psi _ + } \right\rangle \left\langle {\Psi _ + }
\right|,\left| {\Psi _ + } \right\rangle = \frac{1}{{\sqrt 3
}}\left( {\left| {00} \right\rangle  + \left| {11} \right\rangle  +
\left| {22} \right\rangle } \right), \rho _ +   = \frac{1}{3}\left(
{\left| {01} \right\rangle \left\langle {01} \right| + \left| {12}
\right\rangle \left\langle {12} \right| + \left| {20} \right\rangle
\left\langle {20} \right|} \right), \rho _ -   = \frac{1}{3}\left(
{\left| {10} \right\rangle \left\langle {10} \right| + \left| {21}
\right\rangle \left\langle {21} \right| + \left| {02} \right\rangle
\left\langle {02} \right|} \right).$

Evidently, the states $\rho '\left( 0 \right)$ and $ \rho \left( 0
\right)$ are equally useful quantum resources without decoherence.
In Ref.\cite{Horodecki:1999}, Horodecki demonstrated that $\rho'
\left( 0 \right)$ is a free entangled state for $4 < \alpha \le 5$.
It will evolve into $\rho '\left( t \right)$ according to Eq.(1),
whose partial transpose always has a negative eigenvalue in finite
time $t$. So $\rho '\left( t \right)$ never experiences ESD when
subjecting only to the local dephasing noise in Eq.(1). Furthermore,
the state $ \rho' \left( t \right)$ does not undergo DSD in finite
time. To see this, we perform the local operation $ \left( {\left| 1
\right\rangle \left\langle 1 \right| + \left| 2 \right\rangle
\left\langle 2 \right|} \right) \otimes \left( {\left| 1
\right\rangle \left\langle 1 \right| + \left| 2 \right\rangle
\left\langle 2 \right|} \right) $ on $ \rho' \left( t \right)$. The
resulting two-qubit state is easily proved to be entangled and hence
distillable in finite time $t$. So the state $ \rho ' \left( t
\right)$ can always be distilled under the local dephasing noise.

By far we have considered DSD in the presence of local dephase
noise. Another useful physical quantity in quantum-information
problems is the fidelity, which measures to what extent the evolved
state is close to the initial one. For concreteness, we study the
states $\rho(t)$ and $\rho '(t)$ by means of the Bures fidelity
\cite{Bures:1969}. The Bures fidelity of states $\rho$ and $\sigma$
is defined as $ F\left( {\rho ,\sigma } \right) = \left[ {tr\left(
{\sqrt {\sqrt \rho \sigma \sqrt \rho  } } \right)} \right]^2$. For
the state $ {\rho \left( t \right)}$, the fidelity is given by

\begin{equation}
F^\rho  \left( t \right) = \left[ {\frac{1}{{21}}\left( {15 + \sqrt
{6e^{ - \Gamma t} \left( {1 + 2e^{\Gamma t}  + \sqrt {1 + 8e^{\Gamma
t} } } \right)} } \right)} \right]^2,
\end{equation}

\noindent while for $ \rho '\left( t \right)$ it reads

\begin{equation}
F^{\rho '} \left( t \right) = \left[ {\frac{1}{{21}}\left( {15 +
\sqrt {18 + 6\sqrt {1 + 8e^{ - 2\Gamma t} } } } \right)} \right]^2.
\end{equation}

\noindent Eq.(5) and Eq.(6) indicate that the fidelities do not
depend on the coefficient $ \alpha$. In Fig.3, we plot $F^\rho
\left( t \right)$ and $F^{\rho '} \left( t \right)$ as the function
of t with $ \Gamma = 1$. Obviously, the fidelity $F^{\rho '} \left(
t \right)$ is always larger than $F^\rho  \left( t \right)$, which
means that $ \rho \left( t \right)$ degrades faster than $ \rho
'\left( t \right)$. In the infinite time limit, the values of
$F^{\rho '} \left( t \right)$ and $F^\rho  \left( t \right)$ will
approach to $ \left[ {\frac{1}{{21}}\left( {15 + 2\sqrt 6 } \right)}
\right]^2$ and $ \left[ {\frac{1}{{21}}\left( {15 + 2\sqrt 3 }
\right)} \right]^2$, respectively. Thus, we have shown that $ \rho
'\left( t \right)$ have the advantages over $ \rho \left( t \right)$
in two aspects: Firstly, it does not experience DSD in finite time.
Secondly, its fidelity decays slower than that of $ \rho \left( t
\right)$ for all time.

\begin{figure}[ptb]
\includegraphics[scale=0.8,angle=0]{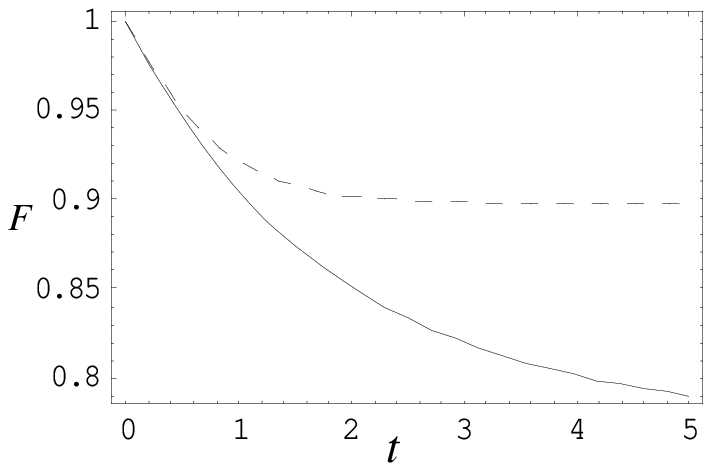}\caption{The fidelity versus time $t$ for different initial states.
The solid and dashed lines correspond to $F^\rho  \left( t \right)$
and $F^{\rho '} \left( t \right)$ with $\Gamma = 1$, respectively.}
\label{fig3}%
\end{figure}

\section{Construction of DSD-free states}

In the last section we have presented the examples of qutrit-qutrit
states with and without DSD. We call the latter as DSD-free states,
which always have negative partial transposition (NPT). NPT is a
necessary condition for distillability under LOCC, so the DSD-free
states are available resource for quantum-information tasks as they
have robust entanglement against local dephasing noise described in
Eq.(1). It is thus worth finding  out diverse DSD-free states in
terms of different kinds of local dephasing noises.
In this sense, DSD-free states deserve storage in
quantum-information "warehouse" and they are more useful than DSD
states.
On the other hand, one can also directly propose schemes of quantum
information processing based on the robustness of DSD-free states.


We now turn to develop several approaches to construct DSD-free
states for an arbitrary qutrit-qutrit state $\sigma$. As a direct
method, we can locally project a few copies of the state $\sigma
(t)$ onto some $2\otimes N (N\geq2)$ states. Such states are
distillable (DSD-free) under LOCC, if and only if they have NPT in
finite time \cite{Horodecki:1997}. Generally, it's difficult to find
out the projectors and there have been a few results on this problem
\cite{Song:2009,Horodecki:1999-2,Chen:2000}. Here we only
investigate one copy of $\sigma (t)$ under the dephasing noise
described in Eq.(1). For example:

\textbf{Lemma 1}. Consider an NPT entangled state $\sigma(0)$ and it
experiences the local dephasing noise in Eq.(1). Then $\sigma(t)$ is
always distillable in finite time, when $\sum_{i= 1}^{2}
D_{2}^{\dagger}\left(t\right)E_{i}^{\dagger}\left(t\right)
\sigma\left(0\right)
E_{i}\left(t\right)D_{2}\left(t\right)\label{krausGeneral}$ or
$\sum_{i= 1}^{2}
D_{i}^{\dagger}\left(t\right)E_{2}^{\dagger}\left(t\right)
\sigma\left(0\right)
E_{2}\left(t\right)D_{i}\left(t\right)\label{krausGeneral}$ is
entangled.

\textit{Proof}. One can get the above given state by projecting
$\sigma(t)$ with the projector $D_{2}\left(t\right)$ on system $A$
or $E_{2}\left(t\right)$ on system $B$ respectively (both projectors
actually have nothing to do with time $t$ when separately used as
operators). Since the given state is a distillable qubit-qutrit
entangled state, the assertion follows from \cite{Horodecki:1997}.
It's interesting to note that the given state is just the sum of the
third and fourth terms in Eq.(1). \hspace*{\fill}$\blacksquare$

Lemma 1 actually requires to check the PPT of some $6 \times 6$
matrices depending on time $t$, which could be done by using
Mathematics. This is a systematic way to build the subspace of
robust entangled states that never experience DSD in Eq. (1). To
show the connection between the last section and lemma 1, we propose
a weak version of lemma 1 as follows.

\textbf{Lemma 2}. Suppose
$D_{2}^{\dagger}\left(t\right)E_{2}^{\dagger}\left(t\right)
\sigma\left(0\right) E_{2}\left(t\right)D_{2}\left(t\right)$ are
entangled, then the states $\sigma(0)$ are DSD-free states in local
dephasing noise Eq.(1).

\textit{Proof}. The state $\sum_{i= 1}^{2}
D_{2}^{\dagger}\left(t\right)E_{i}^{\dagger}\left(t\right)
\sigma\left(0\right)
E_{i}\left(t\right)D_{2}\left(t\right)\label{krausGeneral}$ is
entangled when
$D_{2}^{\dagger}\left(t\right)E_{2}^{\dagger}\left(t\right)
\sigma\left(0\right) E_{2}\left(t\right)D_{2}\left(t\right)$ is
entangled. The assertion then follows from lemma 1.
\hspace*{\fill}$\blacksquare$

One may see that the state $\rho '(0)$ in the previous section
belongs to the case of lemma 2. The advantage of lemma 2 is that the
density operators
$D_{2}^{\dagger}\left(t\right)E_{2}^{\dagger}\left(t\right)
\sigma\left(0\right) E_{2}\left(t\right)D_{2}\left(t\right)$ do not
contain time $t$ and are convenient for calculation. This result is
reasonable for the local dephasing noise described in Eq.(1) which
occurs only between the ground state $i=0$ and the $i$th ($i=1,2$)
excited state.

The model in Eq.(1) is neither the simplest case of local dephasing
nor the most general case. In the most general case dephasing occurs
between all local basis states within each subsystem. In what
follows we put forward some results under such general local
dephasing noise \cite{Ann:2007}.

\textbf{Lemma 3}. Qutrit-qutrit entangled MC states are DSD-free
states in general local dephasing noise.

\textit{Proof}. Let us consider the
qutrit-qutrit maximally correlated (MC) states $\sigma_{MC}(0)$,
which read \cite {Horodecki:2003}
\begin{equation}
\sigma_{MC}(0)=\sum^{2}_{i,j=0}a_{ij}\left|ii\rangle\langle
jj\right|.
\end{equation}
We make it go through the channel of general local dephasing noise.
As we know, such noise cannot change the diagonal elements and the
non-diagonal elements of the evolved state have non-zero dephasing
values for any finite time. So the evolved state $\sigma_{MC}(t)$ is
still a MC state. On the other hand, the MC state is separable if
and only if all non-diagonal entries are zero. Then $\sigma_{MC}(t)$
is always entangled for all finite $t$ whenever the initial state
$\sigma_{MC}(0)$ is entangled. Furthermore, one may prove that
$\sigma_{MC}(t)$ can always be distillable by using the similar
argument to show the distillability of the state $ \rho' \left( t
\right)$ in the previous section.\hspace*{\fill}$\blacksquare$

This example enlightens us on building DSD-free entangled states in
a stronger way, i.e.,

\textbf{Lemma 4}. We perform the local operation $ \left( {\left| i
\right\rangle \left\langle i \right| + \left| j \right\rangle
\left\langle j \right|} \right) \otimes \left( {\left| m
\right\rangle \left\langle m \right| + \left| n \right\rangle
\left\langle n \right|} \right)$ on any initial qutrit-qutrit state,
where $ i,j,m$, and $n$ are chosen from the basis 0, 1 and 2,
respectively. If the projected state is an entangled MC state, then
the initial state is a DSD-free state in general local dephasing
noise.\hspace*{\fill}$\blacksquare$

One can prove lemma 4 by showing that there are always some nonzero
non-diagonal entries of the projected state, and we can take it as
the new initial state in the local dephasing noise. Evidently, the
newly built state in lemma 4 contains MC states in lemma 3.

These results can be extended to the scenario of higher dimension.
For simplicity, we consider the general MC state
$\sigma^{\prime}_{MC}(0)=\sum^{d-1}_{i,j=0}a_{ij}\left|ii\rangle\langle
jj\right|$. Suppose it experiences generalized local dephasing
noise. Analogous to the arguments for lemma 3, one can see that the
state $\sigma^{\prime}_{MC}(t)$ also has no DSD in finite time. Thus
we have provided a high dimensional subspace
$\{|00\rangle,|11\rangle,,...,|d-1,d-1\rangle\}$ in which all
entangled states are distillable under local dephasing noise; that
is,

\textbf{Lemma 5}. Entangled MC states are DSD-free states in general
local dephasing noise. \hspace*{\fill}$\blacksquare$

One can also build other higher dimensional DSD-free states, which
could be locally projected onto entangled MC states by following the
techniques for lemma 4. As a short summary, lemma 1 to 5 give a few
primary results on DSD-free states in finite time, in terms of the
special dephasing noise Eq.(1) and general dephasing noise
respectively.

To gain a better understanding of the entanglement dynamics, we
further study the properties of the evolved state in the
infinite-time limit, which is summarized as the following lemma.

\textbf{Lemma 6}. For any qutrit-qutrit state under the dephasing
noise in Eq.(1), the final evolved state could be separable or
eternally distillable in the infinite-time limit, but no PPT
entangled sate can exist as time goes to infinity.

\textit{Proof}. Let us briefly justify the above statement. Consider
an arbitrary qutrit-qutrit state $\sigma$ under the action of an
infinite time of local dephasing, then the final state $ \sigma
\left( t \right)$ can always be written as: $ a\left| {00}
\right\rangle \left\langle {00} \right| + \left| 0 \right\rangle
\left\langle 0 \right| \otimes \left( {b\left| 1 \right\rangle
\left\langle 1 \right| + c\left| 1 \right\rangle \left\langle 2
\right| + c^* \left| 2 \right\rangle \left\langle 1 \right| +
d\left| 2 \right\rangle \left\langle 2 \right|} \right) + \left(
{e\left| 1 \right\rangle \left\langle 1 \right| + f\left| 1
\right\rangle \left\langle 2 \right| + f^* \left| 2 \right\rangle
\left\langle 1 \right| + g\left| 2 \right\rangle \left\langle 2
\right|} \right) \otimes \left| 0 \right\rangle \left\langle 0
\right| + \left( {\left| 1 \right\rangle \left\langle 1 \right| +
\left| 2 \right\rangle \left\langle 2 \right|} \right) \otimes
\left( {\left| 1 \right\rangle \left\langle 1 \right| + \left| 2
\right\rangle \left\langle 2 \right|} \right)\rho \left( 0
\right)\left( {\left| 1 \right\rangle \left\langle 1 \right| +
\left| 2 \right\rangle \left\langle 2 \right|} \right) \otimes
\left( {\left| 1 \right\rangle \left\langle 1 \right| + \left| 2
\right\rangle \left\langle 2 \right|} \right) $. There are four
terms in all, and each of them corresponds to one of the terms in
Eq.(1), respectively. Note that the first three terms could be
removed by the local operator $ \left( {\left| 1 \right\rangle
\left\langle 1 \right| + \left| 2 \right\rangle \left\langle 2
\right|} \right) \otimes \left( {\left| 1 \right\rangle \left\langle
1 \right| + \left| 2 \right\rangle \left\langle 2 \right|} \right)$
and it does not change the fourth term. So the evolved state $
\sigma \left( t \right)$ is entangled if and only if the fourth term
is entangled. It immediately implies that the state $ \sigma \left(
t \right)$ could be separable or eternally NPT and distillable in
the infinite time limit. As we have seen in the previous sections,
both cases also exist in finite time's evolution. Since the
entanglement property of the evolved state is determined by the
two-qubit state, then we conclude that any PPT entangled state is
unstable as time goes to infinity under the local dephasing noise in
Eq.(1). In contrast, it's still an open question whether there
always exists PPT entangled state in finite time's evolution.
\hspace*{\fill}$\blacksquare$

\section{Discussions and Conclusions}

Before concluding we would like to discuss some open questions as
follow:

Firstly, the initial state in our scheme is a free entangled state.
We have shown that it may evolve into a PPT entangled state and then
a separable state for finite time; it may also be eternally free
entangled for all finite time in the local dephasing noise described
by Eq.(1). We have also shown that any qutrit-qutrit PPT entangled
state will lose its entanglement as time goes to infinity in our
decoherence model. Then an open question is whether PPT entanglement
can be preserved in any finite time under (general) local dephasing
noise? If the answer is no (supported by the calculation in Sec.II),
then we have another interesting difference between the free and
bound entangled state: that is, the former can be robust against the
decoherence in any time while the latter cannot.

Secondly, we have proposed a few systematic methods to build
DSD-free subspaces for $ 3 \otimes 3$ state, in which all states
keep their distillability asymptotically in local dephasing noise.
The primary extension to higher dimensional bipartite system has
also been proposed. It's then an open question to find out more
entangled states which are robust against different local noises
(beyond our model) in any dimensional space. Such states are
applicable to quantum information schemes and thus deserve a deeper
study.

In summary, we have introduced the concept of DSD through an
explicit qutrit-qutrit state. We found that under the bi-local
dephasing noise, the free entangled state may evolve into a PPT
entangled state (DSD) and then become separable (ESD) permanently.
Moreover, a simple local unitary operation on the initial state can
avoid the sudden death of distillability. We also have compared the
action of DSD and DSD-free states in terms of the Bures fidelity.
Next, we have proposed the systematic methods of building the
DSD-free space against decoherence, so that the evolved states are
distillable in finite time. Finally, we proved that there is no PPT
entangled state in infinite-time limit in our model. Our results
imply that further study on free entanglement with time evolution in
practical local noises is required.

\section{Acknowledgements}

We thank A. Winter and C. Beny for critical reading the manuscript.
W. Song was supported by the Doctoral Startup Foundation of
Guangdong Province. The Centre for Quantum Technologies is funded by
the Singapore Ministry of Education and the National Research
Foundation as part of the Research Centres of Excellence programme.
SLZ was supported by NSFC under Grant No. 10674049 and the State Key
Program for Basic Research of China(Nos.2006CB921801 and
2007CB925204).

\end{document}